\documentclass[conference]{IEEEtran}
\IEEEoverridecommandlockouts
\usepackage{cite}
\usepackage{amsmath,amssymb,amsfonts}
\usepackage{algorithmic}
\usepackage{graphicx}
\usepackage{textcomp}
\usepackage{xcolor}
\usepackage{booktabs}

\usepackage{multirow}
\usepackage{pdfpages}
\usepackage{listings}
\usepackage{mwe}
\usepackage{soul}
\usepackage{comment}
\usepackage{svg}
\usepackage{enumitem}
\usepackage{tabu}
\usepackage{makecell}
\usepackage{adjustbox}
\usepackage{tikz}
\usepackage{pgfplots}
\usepackage{balance}
\usepackage{color}
\usepackage{colortbl}
\usepackage{hhline}
\usepackage{multirow}
\usepackage{booktabs}
\usepackage{makecell}
\usepackage[absolute]{textpos}

\usepackage{minted}
\usepackage[ruled,vlined]{algorithm2e}

\definecolor{dscolor}{HTML}{e31a1c}
\definecolor{dsscolor}{HTML}{33a02c}
\definecolor{sscolor}{HTML}{1f78b4}
\definecolor{distcolor}{HTML}{000000}
\definecolor{light-gray}{gray}{0.92}

\definecolor{frcolor}{HTML}{FF0000}

\pgfplotsset{compat = 1.3}

\def\baselinehls baselinehls{Baseline HLS} 
\def\pragmaopt pragmaopt{Directives} 
\def\ourmethod \ourmethod{Optimized}

\floatname{algorithm}{Procedure}

\def\BibTeX{{\rm B\kern-.05em{\sc i\kern-.025em b}\kern-.08em
    T\kern-.1667em\lower.7ex\hbox{E}\kern-.125emX}}
\begin{document}

\title{\fontsize{17}{22}\selectfont
Enabling Heterogeneous Performance Analysis for Scientific Workloads\\
}

\author{\fontsize{18}{22}\selectfont
    \IEEEauthorblockN{Maksymilian Graczyk\IEEEauthorrefmark{1}, Vincent Desbiolles\IEEEauthorrefmark{3}, Stefan Roiser\IEEEauthorrefmark{1}, and Andrea Guerrieri\IEEEauthorrefmark{3}}
    \IEEEauthorblockA{\IEEEauthorrefmark{1}CERN, Geneva, Switzerland}
    \IEEEauthorblockA{\IEEEauthorrefmark{3}School of Engineering, HES-SO Valais-Wallis, Sion, Switzerland}
    \thanks{\textcopyright 2025 IEEE. Personal use of this material is permitted.  Permission from IEEE must be obtained for all other uses, in any current or future media, including reprinting/republishing this material for advertising or promotional purposes, creating new collective works, for resale or redistribution to servers or lists, or reuse of any copyrighted component of this work in other works.}
}

\maketitle

\begin{abstract}
Heterogeneous computing integrates diverse processing elements, such as CPUs, GPUs, and FPGAs, within a single system, aiming to leverage the strengths of each architecture to optimize performance and energy consumption. In this context, efficient performance analysis plays a critical role in determining the most suitable platform for dispatching tasks, ensuring that workloads are allocated to the processing units where they can execute most effectively. \textit{Adaptyst}~\cite{adaptyst} is a novel ongoing effort at CERN,  with the aim to develop an open-source, architecture-agnostic performance analysis for scientific workloads.
This study explores the performance and implementation complexity of two built-in eBPF-based methods such as Uprobes and USDT, with the aim of outlining a roadmap for future integration into \textit{Adaptyst} and advancing toward heterogeneous performance analysis capabilities.

\end{abstract}

\section{Introduction}
In recent years,  heterogeneous platforms enable optimized execution of workloads according to their computational characteristics.
However, the inherent complexity of heterogeneous systems poses significant challenges for performance optimization. Efficient utilization of resources requires a deep understanding of workload behavior across different components. This necessitates robust profiling methods capable of capturing fine-grained performance metrics, identifying bottlenecks, and guiding optimization strategies. Profiling techniques, ranging from hardware counters and sampling-based tools to trace-based analysis, play a critical role in bridging the gap between system potential and actual performance.

\begin{figure}[h]
\centering
\includegraphics[width=1\linewidth]{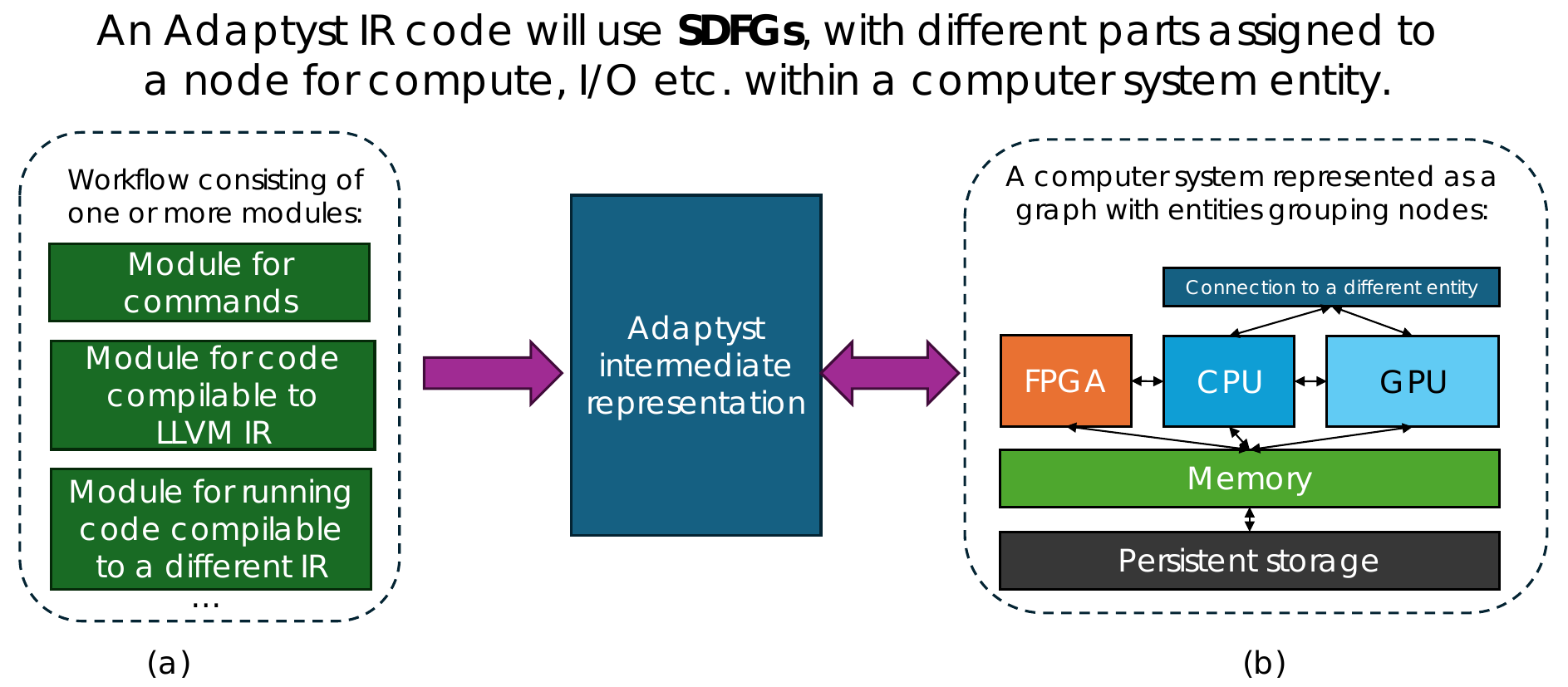}
\vspace{-0.3cm}
\caption{The modular design of \textit{Adaptyst} as a software-hardware co-design framework. Stateful dataflow multigraphs \cite{dace} are used as the IR. (a) Every module describes how a corresponding \textit{Adaptyst} IR should be generated for its use case. (b) Every node is assigned to a backend module which describes how a specific system component should be modelled/profiled, how it can connect to other nodes, and (later) how a match between the component and the \textit{Adaptyst} IR region should be calculated.
}
\vspace{-0.3cm}
\label{adaptyst-future}
\end{figure}

\textit{Adaptyst}~\cite{adaptyst} is a novel, open-source performance analysis and software-hardware co-design tool designed to scale from embedded to high-performance computing and incorporate several static and dynamic analysis methods, including sampling-based profiling. Its modular nature as shown in Fig. \ref{adaptyst-future} and integration with stateful dataflow multigraphs \cite{dace} make our work easily adaptable to already-existing and new solutions on both the software and hardware sides while considering the big picture of computing systems like memory hierarchies, networking, operating systems, and storage. \textit{Adaptyst} can profile on-CPU and off-CPU activity of single- and multi-threaded codes in any compiled language running on Linux. This is made possible through an internal module named \textit{linuxperf} (based on a customized version of \textit{perf}) capable of inspecting CPU performance of arbitrarily-sized code blocks of a program using the cache-aware roofline modelling method. \textit{Adaptyst} has been successfully tested on x86-64, arm64, and RISC-V instruction set architectures.

\section{Related Work and Contribution}
\label{sect:relatedwork}

The profiling ecosystem includes several state-of-the-art tools \cite{perf}, \cite{vtune}, \cite{uprof}, \cite{gprof}, \cite{shende2006tau}, \cite{adhianto2010hpctoolkit}, but few offer extensible support for heterogeneous or non-standard architectures. Additionally, not all are open-source or vendor-portable, limiting their broader applicability. In software-hardware co-design, there are several works targeting different parts of the software-hardware spectrum. This list include standard, modular or transparent compilers \cite{lattner2021mlir}, \cite{dace}, domain-specific languages or vendor-specific programming models \cite{auerbach2012compiler}, \cite{banerjee2000matlab}, \cite{grosser2016polly}, \cite{riebler2019transparent}, or high-level synthesis tools.

\textit{Adaptyst} is designed to interact with existing tools rather than compete with them: our work aims to bridge the gaps between these tools by ensuring language-agnosticism, automatic hardware and code selection, and considering the broader picture beyond solely compute units. However, short-term \textit{Adaptyst}'s development faces some challenges: (1) in the case of \textit{linuxperf}, the necessity of calling a separate instance of "perf" for profiling CPU activity of codes instead of built-in kernel methods like \textit{perf\_event\_open} or eBPF, (2) the requirement of compiling programs with frame pointers for complete stack traces.

\label{sect:performanceanalysis}
In this work, we focus on addressing the first limitation of \textit{Adaptyst} by first investigating the preliminary performance and implementation complexity of two eBPF-based built-in alternatives: Uprobes and USDT (User-Static Defined Tracing). 

\textit{Adaptyst} uses three types of profiling: sampling-based CPU walltime, tracing of context switches for off-CPU activity analysis, and tracing of spawning and exiting threads/processes of a given program. The aforementioned eBPF alternatives (Uprobes and USDT) are related to the latter type, and they may also be expanded later when looking into removing the frame pointer compilation requirement.

\section{Preliminary Results}
\label{sect:finalresults}
For our eBPF investigation, we want to evaluate two primary aspects: (1) performance overhead and stability,  (2) implementation complexity. The first aspect, performance overhead, involves measuring and comparing the runtime overhead introduced by each profiling tool during program execution. This allows for a detailed understanding of the efficiency of each tool and its suitability for real-time or high-performance applications. The second aspect, implementation complexity, focuses on assessing the practical challenges associated with deploying each tool.

To evaluate the profiling methods, a lightweight yet sufficiently complex C program that computes approximate square roots of integers from 1 to 100 has been used. This allows a fast-executing workload for serial benchmarking (multiple runs to average measures), to reveal differences between profiling methods. Compilation was performed with GCC version 13.3.1 using the flags \texttt{-fno-omit-frame-pointer} and \texttt{-mno-omit-leaf-frame-pointer}, ensuring consistent stack traces. The experiments were conducted on an Intel Xeon Silver 4216 with 32 cores, 180 GB RAM, running Gentoo Linux. To minimize scheduling noise, each program was pinned to a dedicated CPU core. 

\subsection{Performance Overhead and Stability}
\begin{figure}
	\centering
    \frame{\includegraphics[width=0.8\columnwidth]{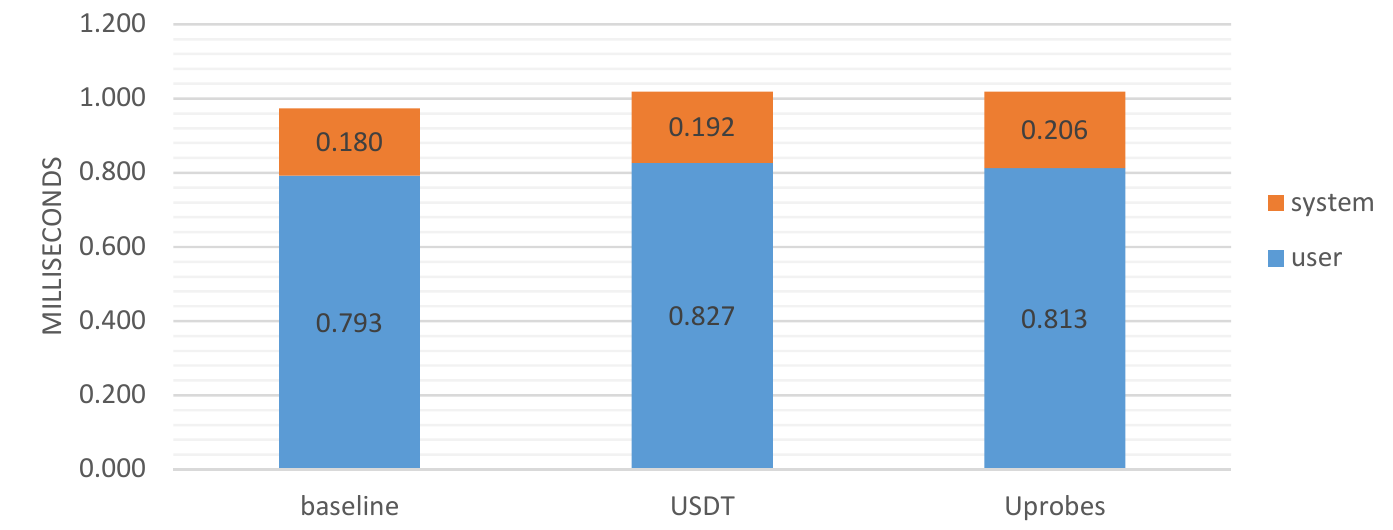}}
	\caption{\small System versus User Breakdown over 100 warm-up runs and 1000 measurement runs. Uprobes uses more system time than USDT.}
	\label{fig:results}
 \vspace{-0.4cm}
\end{figure}

Profiling results (presented in Table \ref{tab:results}) show minimal overhead across all configurations, respectively 5.1\% for USDT and 4.8\% for Uprobes. The standard deviation is modest across all cases, indicating stable performance. While the overall distribution is consistent, the USDT configuration shows slightly higher variance and maximum execution time, suggesting a marginal increase in profiling overhead. Figure \ref{fig:results} shows the breakdown of system and user time for the same configurations. It highlights that Uprobes incurs more system time than USDT or baseline, likely due to kernel-level instrumentation.

\begin{table}[h]
\centering
\caption{Timing results for different profiling methods. Benchmarking was conducted using \textit{hyperfine}~\cite{hyperfine}, with 100 warm-up runs and 1000 measurement runs. Results are expressed in milliseconds.}
\resizebox{\columnwidth}{!}{%
\begin{tabular}{lccccc}
\toprule
\textbf{Type} & \textbf{Mean} & \textbf{Stddev} & \textbf{Median} & \textbf{Min} & \textbf{Max} \\
& \textbf{ (ms) } & \textbf{(ms)} & \textbf{(ms)} & \textbf{(ms)} & \textbf{(ms)} \\
\midrule
Baseline & 1.026 & 0.199 & 0.963 & 0.827 & 1.949 \\
USDT     & 1.079 & 0.211 & 1.004 & 0.839 & 2.083 \\
Uprobes  & 1.076 & 0.207 & 1.000 & 0.833 & 2.054 \\
\bottomrule
\end{tabular}
}
\label{tab:results}
\end{table}

\subsection{Implementation Complexity}
In terms of deployment complexity, eBPF can run concurrently with existing applications without code changes. However, developing eBPF-based solutions involves significant complexity due to intricate data structures and a multi-stage compilation process.

\section{Conclusion and Future Work}
This study presented a preliminary analysis of two eBPF-based profiling tools focusing on operational overhead and deployment complexity. The results show that both USDT and Uprobes introduce minimal overhead compared to the baseline. While all methods demonstrate consistent performance, Uprobes contributes slightly more system time, and USDT exhibits marginally higher variability. Future work includes the integration of e-BPF into \textit{Adaptyst}, as well as the support for non-CPU devices such as GPUs and FPGAs.

\section*{Acknowledgements}
This work is funded by European Union’s Horizon Europe Programme grant 101092877 (SYCLOPS), 101129744 (EVERSE - HORIZON-INFRA-2023-EOSC-01-02), and HES-SO.

\bibliographystyle{IEEEtran}
\bibliography{biblio}

\end{document}